\documentclass[twoside]{article}
\usepackage{fleqn,espcrc2}

\usepackage{graphicx}
\usepackage[figuresright]{rotating}

\newcommand{\AmS}{{\protect\the\textfont2
  A\kern-.1667em\lower.5ex\hbox{M}\kern-.125emS}}

\title{The physics of the stripe quantum critical point in 
the superconducting cuprates}

\author{C. Di Castro, L. Benfatto, S. Caprara, 
C. Castellani, and M. Grilli\address{INFM and Dipartimento di Fisica
Universit\`a di Roma ``La Sapienza'' \\ 
        00185 Rome, Italy}}

\begin{document}

\begin{abstract}

We elaborate on several observable consequences of the Quantum-Critical-Point 
scenario. In particular we show that
the strong k-dependent scattering of the quasiparticles with the 
quasi-critical charge and spin fluctuations reproduces the main features of 
the low-energy spectral weights and  of the observed Fermi surfaces. In the 
underdoped cuprates the attractive k-dependent charge scattering drives the 
formation of the pseudogap at the M points below the crossover temperature 
$T^*$. In this context we discuss models for pseudogap formation with 
relevant scattering  in the particle-particle and particle-hole channels. The 
experimental consequences for the pair-fluctuation and for the 
pseudogap behavior are investigated. 
\vspace{1pc}
\end{abstract}

\maketitle

\section{THE STRIPE QUANTUM CRITICAL POINT SCENARIO}

The non-Fermi-liquid behavior of the normal-state of the  cuprates has two 
major features depending on the doping $(\delta)$ regimes. Specifically, 
(i) near optimal doping no energy scales seem to be present besides the 
temperature (e.g. the in-plane resistivity stays linear in $T$ from just 
above  the critical temperature $T_c$), while (ii) in the underdoped regime, 
new energy 
scales appear in the form of pseudogaps, which persist well above $T_c$ up to 
a doping-dependent crossover temperature $T^*$ \cite{timusk}. Starting in the 
deeply underdoped phase, $T_c$ increases with increasing doping and  $T^*$ 
decreases from high values of several hundreds of kelvins until it merges 
with $T_c$ near optimal doping. On the other hand, a Fermi-liquid-like 
behavior is observed in the overdoped materials. Correspondingly, 
many different physical quantities display qualitatively different 
behaviors in going from the under- to the optimally and to the over-doped 
regimes. As schematically described in Fig. 1, the subdivision of the phase 
diagram in three regions naturally arises from the occurrence of an 
instability line starting at high temperature in the deeply underdoped phase
and ending at zero temperature in a quantum critical point (QCP) located near 
optimal doping. 
In this scheme the optimally doped and overdoped regimes would be related to 
the quantum critical (QC) and to the quantum disordered (QD) region of the 
QCP respectively. The two regions are separated by a crossover line 
$\tilde{T}(\delta)$. The underdoped regime corresponds to the (quasi)-ordered 
region below the instability line. However, precursor effects of the ordering 
could extend up to a higher temperature $T_0(\delta)$.

\begin{figure}
\begin{center}
\includegraphics[width=12pc,height=15pc,angle=-90]{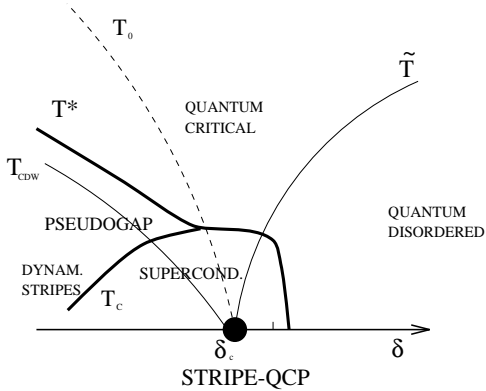}
\caption{Schematic structure of the temperature vs. doping $\delta$ phase 
diagram around the Stripe-QCP.}
\label{fsmodel}
\end{center}
\end{figure}

It was shown that in strongly correlated systems (e.g., in the 
large-U Hubbard model with an 
electron-phonon interaction and long-range Coulomb forces) an incommensurate 
charge-density-wave instability occurs when the doping is reduced below a 
critical value $\delta_c$ \cite{prl95,prb96,jpcs98}. This tendency to order 
the charge arises as a compromise between the local tendency towards phase 
separation and the electrostatic cost to segregate charged carriers \cite{PS}. 
For reasonable values of the parameters this 
instability line starts near optimal doping at zero temperature. In the 
underdoped regime this charge ordering tendency occurs below a 
doping-dependent $T_{CDW}(\delta)$ instability line. The charge ordering 
strongly mixes with spin degrees of freedom and gives rise to the so-called 
stripe phase. 

As shown in Ref. \cite{prl95,jpcs98} a crucial consequence of the stripe 
formation 
is the occurrence nearby the  instability of a singular effective interaction,
strongly dependent on momentum, doping, and temperature:
\begin{equation}
\Gamma ({\bf q},\omega) \approx 
\tilde{U} - \frac{V}{\kappa^{2}+
\vert {\bf q} - {\bf q}_c \vert^2
- i\gamma \omega} 
\label{fitgamlr}
\end{equation}
where $\tilde{U}$ is the residual repulsive interaction between the 
quasiparticles, $\gamma$ is a damping parameter, and ${\bf q}_c$ is the 
wavevector of the CDW instability. The crucial parameter 
$\kappa^2=\xi_c^{-2}$ is 
the inverse square of the correlation length of charge
order and provides a measure of the distance from criticality. 
At $T=0$, in the overdoped regime, $\kappa^2$ is linear in the 
doping deviation from the critical concentration, $\kappa^2=a(\delta-
\delta_c)$. In Ref.  \cite{prl95} the instability was found at 
$\delta_c\approx 0.2$, with $q_c \sim 1$. On the other hand, in the QC region 
above $\delta_c$, $\kappa^2\sim T$, according to the behavior of a Gaussian
QCP. In the underdoped regime $\kappa^2$ vanishes approaching the
instability line $T_{CDW}(\delta)$. 

The occurrence of singular interactions near the QCP and near the 
instability line determines the physical properties of the 
cuprates. In particular, the non-Fermi-liquid behavior characteristic of the 
optimally doped materials is a signature of the QCP \cite{prl95}. In the next 
section we report on some spectroscopic
consequences of the strong scattering mediated by charge fluctuations near 
optimal doping. On the other hand in the overdoped region the term 
$\kappa^2=a(\delta-\delta_c)$ reduces the scattering and determines a region 
of Fermi-liquid behavior. In the underdoped compounds, when the instability 
line $T_{CDW}$ is approached, a singular scattering between the quasiparticles 
is again mediated by the charge fluctuations at wavevectors
${\mbox{\bf $q$}} \approx {\mbox{\bf $q$}}_c$. Thus the region near 
$T_{CDW}(\delta)$ is characterized by a strong effective interaction both in 
the particle-particle (p-p) and the particle-hole (p-h) channels, with a new 
doping-dependent energy scale. In both cases a pseudogap is an expected 
outcome, as it will be discussed in Section 3.

\section{SPECTRAL PROPERTIES NEAR OPTIMAL DOPING}

The charge fluctuations couple with spin degrees of freedom since in the 
hole-poor regions the system is locally closer to half-filling where
antiferromagnetic correlations are more pronounced.  Both charge and spin 
fluctuations then mediate a nearly singular scattering between the
quasiparticles, strongly affecting the spectral properties. In order to 
compare the outcomes of this scattering with the ARPES experiments, mostly 
performed on optimally doped Bi2212 \cite{SAINI}, we assumed \cite{CAPRARA} a 
tight-binding model with the band parameters commonly accepted for this 
material. 
The exchange of QC charge fluctuations at wavevectors ${\bf q}_c=\pm 
(0.4\pi,-0.4\pi)$, and QC antiferromagnetic spin fluctuations at  ${\bf q}_s=
(\pi,\pi)$  was then considered within a perturbative approach. Since in 
general the critical wavevector is model 
and doping dependent, the present choice of ${\bf q}_c$ was suggested
to match the experiments \cite{SAINI}. 
The resulting single-particle spectra are 
characterized by (i) a transfer of spectral weight from the quasiparticle 
peak to the incoherent shadow peaks; (ii) a redistribution of the
low-energy spectral weight with a modification of the FS; (iii) a strong 
anisotropic suppression of spectral weight around the M points
$(\pm\pi,0)$ and $(0,\pm\pi)$. All these 
features have a counterpart in the experiments. 

In this framework one can also investigate the bilayer structure of Bi2212 
and explain the puzzling absence of bonding-antibonding band
splitting. According to the band calculations \cite{andersen},
we introduced \cite{CAPRARA2} a k-dependent interplane hopping 
$t_\perp({\bf k})=t_\perp |\gamma_{\bf k}|$
with $\gamma_{\bf k}={1\over 2}(\cos k_x -\cos k_y )$. 
$t_\perp({\bf k})$ is large near the M points only. The intraplane scattering, 
however, mostly reduces the quasiparticle spectral weight near the M points.
Thus, within our scenario, the absence of detectable band splitting in ARPES 
spectra follows directly from the in-plane spectral properties.

The interaction mediated by the quasicritical fluctuations also provides an 
effective pairing mechanism. In this 
regard approaching the superconducting region from the overdoped regime, the 
doping and the temperature dependences of the $\kappa^2$ term in the QD and QC 
regimes give rise to a non-trivial increase of $T_c$ followed by a 
saturation around optimal doping \cite{prb96}. The more involved case of the 
underdoped regime, with pseudogap formation will be discussed in the next 
section.

\section{PARTICLE-PARTICLE AND PAR\-TI\-CLE-HOLE PSEUDOGAP}

The effect of the stripe instability can be more dramatic in underdoped 
materials, when the system approaches the instability line at temperatures 
$T\sim T_{CDW}(\delta)$. In this case, near $T_{CDW}(\delta)$ the critical 
fluctuations at 
wavevectors near ${\bf q}_c$ can mediate a large effective interaction 
between the quasiparticles states ${\bf k}$ and ${\bf k'}$ 
such that ${\bf k}-{\bf k'}\sim {\bf q}_c$. The generic outcome is that 
states of the Fermi surface near the M points are strongly interacting, while 
quasiparticles around the diagonals ($\Gamma-X$ and $\Gamma-Y$ directions) 
are less affected. This finds a correspondence in ARPES experiments 
\cite{marshall,ding}, where at $T^*$ the Leading Edge shift ($LE$) starts to 
develop near the M points. Indeed the strong interaction near the M points 
can give rise to pairing and gaps both in the p-p and  p-h channels. Although 
it is quite natural that both channels contribute to the formation of the 
pseudogap below 
the crossover temperature $T^*$, the two limiting cases, when a single 
channel (either p-p or p-h) dominates the pseudogap formation, are simpler to 
analyze and each one of them shows relevant aspects of the physics of the 
cuprates.

In the first mechanism we propose, the pseudogap opens due to incoherent 
pairing in the p-p channel. The strong momentum dependence of the effective 
interaction (\ref{fitgamlr}) plays in this regard a crucial role in selecting 
the quasiparticle states which are most strongly paired. This leads to 
non-trivial fluctuation effects, because strongly paired states near the M 
points coexist with weakly interacting quasiparticles along the diagonals. 
This situation is quite different from the case described by a single 
superconducting order parameter $\Delta ({\bf k})=\Delta_s g({\bf k})$. In 
the present case, indeed, the momentum dependence of the effective pairing 
interaction  not only produces the k structure of $g({\bf k})$, but also 
confers different fluctuation properties  {\em in k-space} to the Cooper 
pairs depending on their strongly or weakly paired character. This physical 
situation has recently been described within a two-gap model \cite{twogap}. 
In this latter framework, incoherent tightly bound Cooper pairs around the M 
points are formed at $T^*$, while phase coherence is established at a lower 
temperature $T_c$ by coupling to the stiffness of the weakly bound pairs near 
the diagonal directions.

\begin{figure}
\begin{center}
\includegraphics[width=12pc,height=15pc,angle=-90]{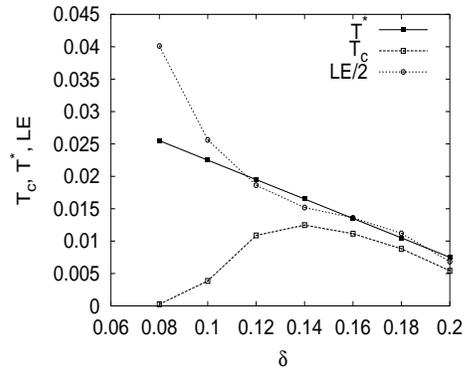}
\caption{Doping dependence of $T^*$, $T_c$, and the zero-temperature
$LE$. We assumed the $T^*(\delta)$ dependence
measured in Bi2212 \cite{ding} and used $c=2.5$ and $V=90$ meV to reproduce
the experimentally observed values of the $LE$ in the underdoped region
and the maximum $T_c$.}
\label{tctstar}
\end{center}
\end{figure}

The scattering in the p-h channel can provide an additional mechanism for the 
pseudogap formation. If this happens, the issue arises of the interplay 
between the preformed p-h pseudogap and an additional BCS pairing for the 
weakly interacting quasiparticles. While this issue was discussed in Ref.
\cite{nozieres} for a simple isotropic pseudogap, in Ref. \cite{benfatto} a 
specific band structure is considered, which includes a preformed k-dependent 
gap. The whole complication of the strong scattering around the M points is 
schematized by this preformed p-h gap $\Delta_0(\delta,T) \gamma_{\bf k}$, 
which separates the conduction and the valence band, and vanishes at the 
points $(\pm \pi/2,\pm \pi/2)$. Each band has a width $4t\simeq 1$ eV, $t$ 
being the nearest-neighbor hopping. We assume $T^*(\delta)$ as the critical
line for the preformed gap formation and take $\Delta_0(\delta,T)=c 
T^*(\delta) g(T/T^*(\delta))$, where $c$ is a fitting parameter, $g(0)=1$,
$g(1)=0$, and $g(x)$ interpolates smoothly between these two limits. A 
suitable weak pairing $V$ in the Cooper channel promotes a d-wave 
superconducting gap $\Delta_s(\delta,T) \gamma_{\bf k}$ in the low valence 
band of the hole doped system. The mean-field BCS critical temperature $T_c$ 
vanishes at $\delta=0$, increases with increasing doping, and reaches a maximum
at $\delta_c$ when the chemical potential crosses the peak which individuates 
the pseudogap region in the density of states, and then decreases. Therefore 
$T^*$ and $T_c$ merge near optimum doping and the $T_c(\delta)$ curve has the 
characteristic bell-shaped form in reasonable agreement with the experiments 
(see Fig. 2).

The quasiparticle spectra are characterized by a $LE$, i.e. a finite minimum
distance of the quasiparticle peak from the Fermi level, which persists in 
the normal state and is largest at the M points, where $LE\simeq \Delta_0-
|\mu|$. In underdoped SC regime, the $LE$ is controlled by two parameters, as 
seen in experiments \cite{PX,TWO}. The M points are dominated by the 
normal-state pseudogap, whereas the nodal region are controlled by 
$\Delta_s$, which scales as $T_c$. In the overdoped regime the $LE\simeq
\Delta_s$.

The above preformed gap accounts for most of the non-mean-field effects by 
the input of a normal-state pseudogap $\Delta_0(\delta,T)$. In particular, 
the model yields a phase diagram in good qualitative agreement with the 
experiments. 

On the other hand, the bifurcation between $T^*$ and $T_c$ nearby optimum 
doping is also an outcome of the two-gap model \cite{twogap}. Clearly, the 
two models assign a different relevance to the effect of the strong QCP 
effective interaction in the p-p and p-h channels, and select just one of the 
two channels as the most affected one. It is quite plausible that the stripe 
fluctuations will indeed produce non-Fermi-liquid- and non-mean-field-like 
effects in both channels. However, whether the results discussed above within 
each of the two models should cooperate to produce a better quantitative 
description of the cuprates, is still an open problem under investigation.

\end{document}